\documentclass[epjST]{svjour}
\usepackage{graphics}
\usepackage{amsfonts}
\usepackage{amsmath}
\usepackage{amssymb}
\usepackage{graphicx}

\begin{document}
\title{Equilibrium and Nonequilibrium Thermodynamics of a Photon Gas in the Near Field}
\author{Agust\'{i}n P\'{e}rez-Madrid\inst{1}\fnmsep\thanks{\email{agustiperezmadrid@ub.edu}} \and Luciano C. Lapas\inst{2}\fnmsep\thanks{\email{luciano.lapas@unila.edu.br}} \and J. Miguel Rubi\inst{1}\fnmsep\thanks{\email{mrubi@ffn.ub.es}}}
\institute{Departament de F\'{i}sica de la Mat\`eria Condensada, Facultat de F\'{i}sica, Universitat de Barcelona, Mart\'{i} i Franqu\`es 1, 08028 Barcelona, Spain. \and Interdisciplinary Center for Natural Sciences, Universidade Federal da Integra\c{c}\~{a}o Latino-Americana, P.O. Box 2067, 85867-970 Foz do Igua\c{c}u, Brazil.}
\abstract{
In this paper we study the near-field thermodynamics of a photon gas at equilibrium as well as out-of-equilibrium in the presence of dissipative effects. As a consequence of Heisenberg's uncertainty principle, we are able to eliminate the low-frequency modes in both cases, providing an analytical expression for the near-field entropy. In addition, we obtain the entropic-force contributions to the Casimir effect. At zero temperature the well-known $l^{-4}$ behavior of the pressure is obtained. In the nonequilibrium case, we compute the entropy production, showing that the excess of heat in each bodies must be dissipated into the respective thermal reservoirs.}

\maketitle

\section{Introduction}
In the near-field regime, when the separation distance between bodies is intermediate between microscopic and macroscopic length scales of the order of nanometers, interesting phenomena such as Casimir-Lifshitz forces and giant radiative heat transfer occurs. However, a problem arises when we deal within a nonequilibrium regime. The most common approach is to use Rytov theory according to which the electromagnetic-field fluctuations are created by fluctuating electric and magnetic polarizations due to thermal agitation of matter. The equations satisfied by these electromagnetic fields were derived by Landau and Lifshitz \cite{Landau84a} using the fluctuation-dissipation theorem established by Callen and Welton \cite{Callen51}.

All of this applies in local thermal equilibrium \cite{Latella14,Latella15,Latella16}, thus making possible the operation of averaging the stress tensor in a medium separating two bodies. In this way one can find a general expression for the electromagnetic stress
tensor for arbitrary time-dependent fields in that medium. Nonetheless, in the presence of dissipation as occurs in systems in nonequilibrium stationary states, the application of the previously mentioned formalism might not be appropriate . Additionally, in the Rytov or fluctuational electrodynamic framework, in order to find the field we usually solve the Maxwell equations by linear response theory \cite{Intravaia11}, which once again might not be an appropriate procedure when dissipation is present since the system can be far from equilibrium.

It is our contention here to propose another scheme which takes into account the non-Markovian character of the system related to dissipative effects and the complexity of the relaxation phenomena which take place there. We will undertake this by postulating kinetic balance equations for the energy taking into consideration memory effects.

\section{Equilibrium Thermodynamics}

It is known that at zero temperature the energy spectrum of vacuum quantum
fluctuations are given by $u(\omega)=\hbar\omega/2$, with $\hbar$ being the reduced
Planck's constant ($\hbar=h/2\pi$), and thus the Helmholtz
free energy of a  gas of photons confined between two parallel metallic plates at zero temperature is given by
\begin{equation}
F=\int_{\omega_{m}}^{\omega_{D}}u(\omega)g(\omega)d\omega,  \label{free energy}
\end{equation}
where $g(\omega)=V\omega^{2}/\pi^{2}c^{3}$ is the distribution of
modes, with $V=l·A$ being the volume of the cavity, $l$ the separation between plates and $A$ a surface area. We have assumed the existence of a maximum frequency  $\omega_{D}=2\pi c/D$, with $D$ being the inter-atomic distance. Moreover, $\omega_{m}=2\pi \eta c/l$ is a phonon-like cutoff frequency, with $c$ being the speed of light in vacuum and $\eta$ a non-dimensional fitting parameter, inferred in some of our previous publications \cite{Lapas16}. This cutoff frequency, which has been derived after a reinterpretation of the Heisenberg's uncertainty principle~\cite{Perez-Madrid13}, separates the low frequency band from the rest of the modes. This low frequency modes are excluded in the near-field. Hence, from Eq. (\ref{free energy}) we derive the force which is given by
\begin{equation}
f=-\frac{\partial F}{\partial l}=-\frac{\hbar\eta \omega _{m}^{3}}{\pi c^{2}l^{2}}V.  \label{force}
\end{equation}
Hence, from Eq. (\ref{force}) 
\begin{equation}
f=-8\pi^2\frac{\hbar \eta^4 c}{l^{4}}A  \label{force2}
\end{equation}
and for the pressure the following scaling law 
\begin{equation}
p\sim -\frac{\pi^2\hbar c}{l^{4}}\label{pressure}
\end{equation}
can be derived.

Thermal corrections to the previous results can be obtained from the general
expression of the free energy 
\begin{equation}
F=U-TS.  \label{free energy2}
\end{equation}
 Here, we must consider $U=U(T)$, with 
\begin{equation}
U(T)=\int_{\omega_m }^{\omega _D}u(\omega ,T)g(\omega )d\omega ,
\label{internal energy}
\end{equation}%
 where now we consider
\begin{equation}
u(\omega ,T)=\hbar \omega \left[ n(\omega ,T)+\frac{1}{2}\right] =
\frac{\hbar \omega }{2}\coth \left( \frac{\hbar \omega }{k_{B}T}\right)  
\label{oscillator energy}
\end{equation}%
and 
\begin{equation}
n(\omega ,T)=\frac{1}{e^{\frac{\hbar \omega }{k_{B}T}}-1}.
\label{planck distribution}
\end{equation}%
In order to compute the entropy, we know that at constant volume 
\begin{equation}
\frac{1}{T}=\frac{dS}{dU}  \label{temperature}
\end{equation}%
and thus, 
\begin{equation}
S(T)=\int_{0}^{T}dt\frac{1}{t}\frac{dU(t)}{dt}.  \label{entropy}
\end{equation}%
Now, by combining Eqs. (\ref{internal energy}) and (\ref{entropy}), we obtain
\begin{equation}
S(T)= k_{B}\int_{\omega _{m}}^{\omega_D }d\omega g(\omega )\int_{0}^{T}dt\frac{%
\hbar \omega }{k_{B}t}\frac{dn(\omega ,t)}{dt}.  \label{entropy2}
\end{equation}%
Since according to Eq. (\ref{planck distribution}), one can write
\begin{equation}
\frac{\hbar \omega }{k_{B}t}=\ln \left[ 1+n(\omega ,t)\right] -\ln n(\omega
,t),  \label{intermediate}
\end{equation}%
which leads to
\begin{eqnarray}
\int_{0}^{T}dt\frac{\hbar \omega }{k_{B}t}\frac{dn(\omega ,t)}{dt}&=&n(\omega
,T)\left\{ \ln \left[ 1+n(\omega ,T)\right] -\ln n(\omega ,T)\right\}+   \notag\\
&&+\int_{0}^{T}dt\frac{d}{dt}\ln \left[ 1+n(\omega ,t)\right]. \label{intermediate2}
\end{eqnarray}%

\begin{figure}[t]
\centering\resizebox{1\textwidth}{!}{%
  \includegraphics{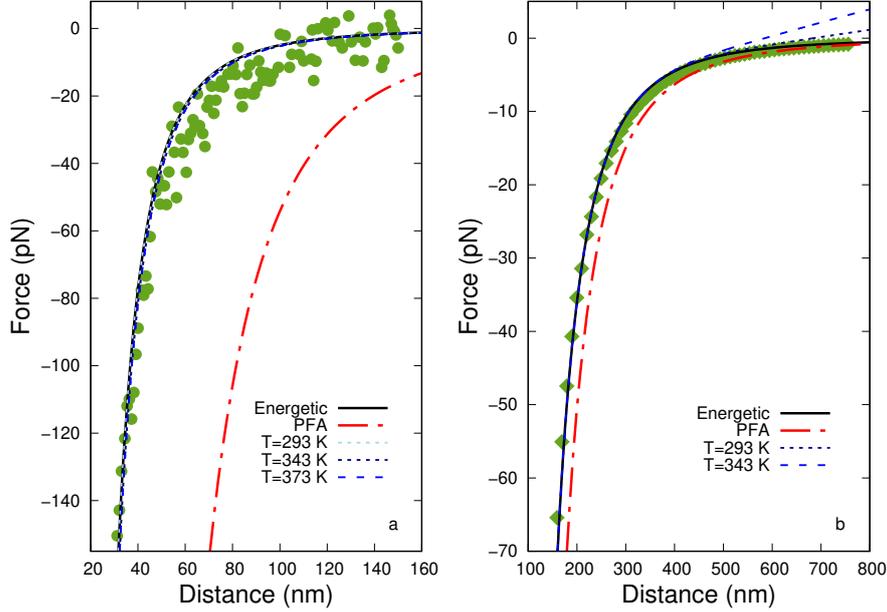}}
  \caption{Attractive Casimir-Lifshitz force comparison. {\bf a.} (Left) Casimir-Lifshitz force between a gold sphere and silica plate versus gap distances for different temperatures. Green circles represent data from Munday {\it et al.} \cite{Munday09,Rodriguez11} using an atomic force microscopy technique for measurement conducted between a large plate and a 39.8 $\mu$m diameter sphere. {\bf b.} (Right) Casimir-Lifshitz force between a gold sphere and gold plate versus gap distances for different temperatures. The green diamonds represent data from Krause {\it et al.} \cite{Krause07}, multiplied by Casimir force assuming that the PFA is valid, using a micro-electro-mechanical torsion oscillator for measurement conducted between a large plate and a 148.2 $\mu$m radius sphere.}
  \label{fig:1}
\end{figure}

Finally, from Eqs. (\ref{entropy2})--(\ref{intermediate2}) we find the
following expression%
\begin{equation}
S=k_{B}\int_{\omega _{m}}^{\omega_D}m(\omega ,T)g(\omega )d\omega ,
\end{equation}%
with%
\begin{equation}
m(\omega ,T)=\left[ 1+n(\omega ,T)\right] \ln \left[ 1+n(\omega ,T)\right]
-n(\omega ,T)\ln n(\omega ,T).  \label{entropy sepectrum}
\end{equation}%
Hence, 
\begin{equation}
\frac{\partial U}{\partial l}=8\pi^{2}V\frac{\hbar c\eta^4}{l^{5}}\coth
\left( \frac{\hbar \pi\eta c}{k_{B}Tl}\right)   \label{energetic force}
\end{equation}%
and  
\begin{equation}
T\frac{\partial S}{\partial l}=8\pi V\frac{k_{B}T}{l^{4}}m(\omega
_{m},T)  \label{entropic contribution}
\end{equation}%
consist in the energetic and entropic contribution to the force which, when
joined together, ultimately lead to the expression of the pressure%
\begin{equation}
p=-8\pi ^{2}\frac{\hbar  c \eta^4}{l^{4}}\coth \left( \frac{\hbar \pi \eta c}{k_{B}Tl}%
\right) +8\pi\frac{k_{B}T}{l^{3}}m(\omega _{m},T).  \label{pressure2}
\end{equation}%
Since it is customary  to relate the pressure to the experimentally determined force gradient along the gap distance, we compare our results with two experiments carried out between a sphere (of radius $R$) and plate. Both of them prove the validity of our approach for attractive Casimir forces observed by Munday {\it et al.}~\cite{Munday09} and by Krause {\it et al}~\cite{Krause07}. In this circumstance, since $R \gg l$, the force was calculated from pressure, Eq. (18), multiplied by an effective area $A=2\pi R l$.  As seen in Fig. \ref{fig:1}, a good agreement is obtained with experimental results, which also are compared with the predictions by the proximity-force approximation (PFA) without corrections (see Ref.~\cite{Krause07} for more details). The attractive behavior displayed in Fig.~\ref{fig:1}{\bf (a)} is obtained with $\eta = 6.40\times 10^{-2}$, whereas in Fig.~\ref{fig:1}{\bf (b)}, $\eta = 1.06\times 10^{-1}$. For short distances, {\it i.e.} $l<400$ nm, energetic contribution (the first term of right hand side in Eq.~\ref{pressure2}) is larger than entropic one. As shown in Fig.~\ref{fig:1}{\bf (b)}, entropic contribution may emphasize for $l>400$ nm. Another approach to the attractive and repulsive Casimir forces can be found in Ref.~\cite{Lapas16}.

Hereinafter, it is possible to analyze some limits: for $l< \lambda _{T}
$ one has 
\begin{equation}
\left[ 1+n(\omega ,T)\right] \ln \left[ 1+n(\omega ,T)\right] \sim 0\text{ ,}
\label{approximations1}
\end{equation}
\begin{equation}
n(\omega ,T)\ln n(\omega ,T)\sim -\frac{\lambda _{T}}{l}e^{-\lambda
_{T}/l} \label{approximations2}
\end{equation}
and
\begin{equation}
\coth \left( \frac{\hbar \pi  \eta c}{k_{B}Tl}\right) \sim 1. \label{approximations3}
\end{equation}
and therefore, we infer 
\begin{equation}
p=-8\pi ^{2}\frac{\hbar \eta^4 c}{l^{4}}+16\pi ^{2}\frac{\hbar c}{l^{4}}e^{-\lambda
_{T}/l},  \label{low temperature}
\end{equation}%
which for extremely low $T$ ($e^{-\lambda
_{T}/l}\approx 0$) coincides with Eq. (\ref{pressure}). On the other hand, if $\lambda _{T}< l$, 
\begin{equation}
\coth \left( \frac{\hbar \pi \eta c}{k_{B}Tl}\right) \sim 2\pi \frac{k_{B}Tl}{%
\hbar c}  \label{approximations4} 
\end{equation}
\begin{equation}
1+n(\omega ,T)\sim n(\omega ,T)\sim \frac{l}{\lambda _{T}}\label{approximations5} 
\end{equation}
and
\begin{equation}
m(\omega _{m},T)\sim 0\label{approximations6}.
\end{equation}
Thereby we obtain the following scaling law%
\begin{equation}
p\sim-\frac{k_{B}T}{l^{3}}.  \label{high temperature}
\end{equation}

We may also compute the density of the photon gas 
\begin{equation}
\rho\left( T\right) =\frac{1}{V}\int_{\omega_{m}}^{\omega_D}n(\omega
,T)g(\omega)d\omega.  \label{density1}
\end{equation}
After performing the change of variable $\omega=1/s$, Eq. (\ref%
{density1}) can be rewritten 
\begin{equation}
\rho\left( T\right) =\frac{1}{V}\int_{{\omega_D}^{-1}}^{\omega_{m}^{-1}}n(\omega\left(
s\right) ,T)g(\omega\left( s\right) )\frac{ds}{s^{2}}.  \label{density2}
\end{equation}
According to the mean value theorem, for very short $l$ ($\omega_{m}$ large)
we can approximate Eq. (\ref{density2}) by multiplying the integrand
computed in an intermediate value $\tilde{s}=\chi\omega_{m}^{-1}$ ($\chi<1$)
by the width of the integration interval $\left(\omega_{m}^{-1}-\omega_{D}^{-1}\right),$ resulting in%
\begin{equation}
\rho\left( T\right) \approx\frac{1}{V}\omega_{m}^{\prime2}\left(\omega_{m}^{-1}-\omega_{D}^{-1}\right)%
n\left( \omega_{m}^{\prime},T\right) g\left( \omega_{m}^{\prime }\right) ,
\label{density3}
\end{equation}
with $\omega_{m}^{\prime}\equiv\omega\left( \tilde{s}\right) =\omega
_{m}/\chi=2\pi\varepsilon c/l$ and where we have introduced the new
parameter $\varepsilon=\eta/\chi$. Therefore, assuming $\left(\omega_{m}^{-1}-\omega_{D}^{-1}\right)\approx \omega_{m}^{-1}$, one has
\begin{equation}
\rho\left( T\right) \approx\frac{8\pi\varepsilon^{4}}{\eta l^{3}}\frac {1}{%
e^{h\varepsilon c/k_{B}Tl}-1}  \label{density4}
\end{equation}
or in \ terms of $\lambda_{T}$ defined above
\begin{equation}
\rho\left( T\right) \approx\frac{8\pi\varepsilon^{4}}{\eta l^{3}}\frac {1}{%
e^{\varepsilon\lambda_{T}/l}-1},  \label{density5}
\end{equation}
an expression with the asymptotic limits:  
\begin{equation}
\rho\left( T\right) \approx\frac{8\pi\varepsilon^{4}}{ \eta l^{3}}%
e^{-\varepsilon\lambda_{T}/l}
\end{equation}\label{density6}
for $l<\lambda_{T}$. In the case $\lambda_{T}< l$, Eq. (\ref{density1}) splits into two terms
\begin{equation}
\rho\left( T\right) =\frac{1}{V}\int_{\omega_{m}}^{\omega_{o}}n(\omega,T)g(\omega)
d\omega +\rho^\prime(T)  \label{density8}
\end{equation}
 and 
\begin{equation}\label{F}
\rho^\prime(T)=\frac{1}{V}\int_{\omega_{o}}^{\omega_{D}}n(\omega
,T)g(\omega)d\omega.
\end{equation}
where $\omega_{o}=k_B T/\hbar$. In this limit, taking into account the chain of inequalities $\omega_m<\omega_o<\omega_D$, we find 
\begin{equation}\label{density9}
\rho\left( T\right) \approx\frac{1\omega_o}{2\pi^2c^3}\left(\omega_o^2-\omega_m^2\right)+\rho^\prime(T)
\end{equation}
and
\begin{equation}\label{F2}
\rho^\prime(T)\approx 2.7\frac{\omega_{o}^{3}}{\pi^2 c^3}
\end{equation}

Note that since the Casimir phenomenon we describe here deals with attractive forces,  the bodies must be pulled apart in order to maintain a constant volume confining the gas of photons which means exerting a negative pressure. Otherwise, increasing the density implies diminishing the pressure and thus bringing about a collapse.

\section{Nonequilibrium thermodynamics}

To study this we will follow a kinetic approach according to which two objects thermalized at different temperatures interchange energy by thermal radiation in such a way that the energy balance equations are%
\begin{equation}
\frac{du_{1}}{dt}=-J_{1\rightarrow2}(t)+J_{2\rightarrow1}(t)+J_{1}(t),
\label{kinetic1}
\end{equation}%
\begin{equation}
\frac{du_{2}}{dt}=-J_{2\rightarrow1}(t)+J_{1\rightarrow2}(t)+J_{2}(t),
\label{kinetic2}
\end{equation}
where $u_{1}$ is the energy of the first object and $u_{2}$ the energy of the second one, both of which are thermalized at temperatures $T_{1}$ and $T_{2}$, respectively. In addition, Eqs. (\ref{kinetic1}) and (\ref{kinetic2}) contain the radiation currents $J_{i\rightarrow j}(t)$ and the heat transferred to the corresponding thermal baths, $J_{1}(t)$ and $J_{2}(t)$. In both bodies mentioned above, the interaction with the radiation brings about transitions among different energy states removing these bodies from equilibrium.

In the stationary state $du_{1}/dt$ and $du_{2}/dt$ vanish and thus,%
\begin{equation}
J_{1}=J_{1\rightarrow 2}-J_{2\rightarrow 1}  \label{kinetic3}
\end{equation}%
and%
\begin{equation}
J_{2}=J_{2\rightarrow 1}-J_{1\rightarrow 2},  \label{kinetic4}
\end{equation}%
implying that $J_{1}=-J_{2}$. Hence, 
\begin{equation}
\dot{S}_{1}=-\frac{J_{1}}{T_{1}}  \label{rate entropy 1}
\end{equation}%
and%
\begin{equation}
\dot{S}_{2}=-\frac{J_{2}}{T_{2}}  \label{rate entropy 2}
\end{equation}%
constitute the constant rate of entropy generation in each bath. Therefore, by combining both Eqs. (\ref{rate entropy 1}) and (\ref{rate entropy 2}) we find 
\begin{equation}
\dot{S}_{1}+\dot{S}_{2}=J_{1}\left( \frac{1}{T_{2}}-\frac{1}{T_{1}}\right) ,
\label{entropy production}
\end{equation}%
which gives us the very important result of the stationary entropy
production of the universe. From here, Onsager's theory enables us to write 
\begin{equation}
J_{1}=X\left( \frac{1}{T_{2}}-\frac{1}{T_{1}}\right) \equiv C\left(
T_{1}-T_{2}\right)   \label{onsager law}
\end{equation}%
where $X$ is a Onsager coefficient relating the thermodynamic current $J(T_{1})$ with the thermodynamic force $\left( 1/T_{2}-1/T_{1}\right) $ and $C=X/(T_{2}T_{1})$ is the thermal conductance which in the general case is a
function of $T_{1}$ and $T_{2}$. However, in the linear case $C=X\left(
T_{0}\right) /T_{0}^{2}$, with $T_{0}=\left( T_{1}+T_{2}\right) /2$.

By bringing back Eq. (\ref{oscillator energy}), one can write
\begin{equation}
J_{2\rightarrow 1}=\frac{u\left( \omega ,T_{2}\right) }{\tau _{2}\left(
\omega ,T_{2}\right) }-\frac{u\left( \omega ,T_{0}\right) }{\tau _{2}\left(
\omega ,T_{0}\right) }  \label{stationary current1}
\end{equation}%
and 
\begin{equation}
J_{1\rightarrow 2}=\frac{u\left( \omega ,T_{1}\right) }{\tau _{1}\left(
\omega ,T_{1}\right) }-\frac{u\left( \omega ,T_{0}\right) }{\tau _{1}\left(
\omega ,T_{0}\right) },  \label{stationary current2}
\end{equation}%
where $\tau _{1}\left( \omega ,T_{1}\right) $ and $\tau _{2}\left( \omega
,T_{2}\right) $ correspond to the spectrum of relaxation times of materials $%
1$ and $2,$ respectively. These relaxation times are related to the
corresponding dielectric constants $\in _{1}\left( \omega \right) $ and $\in
_{2}\left( \omega \right) $  and describe thermal
processes. Assuming 
\begin{equation}
\tau _{\alpha }^{-1}\left( \omega ,T_{\alpha }\right) =\frac{1}{\sqrt{2\pi }%
\sigma _{\alpha }\hat{\tau}_{\alpha }}\exp \left[ -\frac{\ln^{2}\left(
\omega /\omega _{\alpha }\right) }{2\sigma _{\alpha }^{2}}\right] ,
\label{relaxation time}
\end{equation}%
a log-normal spectrum of frequencies which captures the complexity of
thermal relaxation in real materials \cite{Lapas16,Perez-Madrid13,Denisov90,Chumakov11}, where $\hat{\tau}_{\alpha }$ and $\sigma _{\alpha }$ are material-dependent parameters
and $\omega _{\alpha }=k_{B}T_{\alpha }/\hbar $ is the thermal frequency.
Therefore, from Eqs. (\ref{kinetic3}), (\ref{stationary current1}) and (\ref%
{stationary current2}), after linearization, comparison with Eq. (\ref{onsager law}) yields 
\begin{equation}
C\left( \omega \right) =\frac{k_{B}}{\tau \left( \omega \right) }\left[ 
\frac{\hbar \omega /2k_{B}T_{0}}{\sinh \left( \hbar \omega
/2k_{B}T_{0}\right) }\right] ^{2},  \label{conductance}
\end{equation}%
where now $\tau _{\alpha }^{-1}\left( \omega ,T_{\alpha }\right) \approx
\tau _{\alpha }^{-1}\left( \omega ,T_{0}\right) $ and 
\begin{equation}
\frac{1}{\tau \left( \omega \right) }=\frac{1}{\tau _{1}\left( \omega
\right) }+\frac{1}{\tau _{1}\left( \omega \right) } , \label{mathiessen}
\end{equation}%
which corresponds to Mathiessen rule. The conductance given through Eq. (\ref{conductance}) has been successfully
tested by comparison with experiments in one of our previous results
\cite{Perez-Madrid13}.

From Eqs. (\ref{entropy production}), (\ref{onsager law}) and (\ref{conductance}) we can write the expression of the entropy production, $\sigma(\omega)\equiv\dot{S}_{1}+\dot{S}_{2}$  as follows
\begin{equation}
\sigma(\omega)=\frac{k_{B}}{\tau \left( \omega \right) }\left[ 
\frac{\hbar \omega /2k_{B}T_{0}}{\sinh \left( \hbar \omega
/2k_{B}T_{0}\right) }\right] ^{2}\frac{\left(
T_{1}-T_{2}\right)^2}{T_0^2}, \label{sigmaomega}
\end{equation}
which can be integrated to give the entropy production per unit area
\begin{eqnarray}
\Sigma &\equiv& \frac{1}{A}\int_{\omega_{m}}^{\omega_D}\sigma(\omega)g(\omega)d\omega=\frac{1}{A}\int_{0}^{\omega_{m}^{-1}}\sigma(\omega\left(
s\right))g(\omega\left( s\right) )\frac{ds}{s^{2}} \notag \\
&\approx&\frac{1}{A}\frac{\omega_{m}^{\prime2}}{\omega _{m}%
}\sigma\left( \omega_{m}^{\prime}\right) g\left( \omega_{m}^{\prime }\right) \notag \\
&\approx& 2\pi\varepsilon^4\eta \frac{\lambda_{T_0}^2}{l^4} \frac{k_{B}}{\tau \left( \omega_m/\chi \right) }\left[ 
\frac{1}{\sinh \left( \varepsilon\lambda_{T_0}/2l\right) }\right] ^{2}\frac{\left(
T_{1}-T_{2}\right)^2}{T_0^2},\label{Sigmaomega}
\end{eqnarray}
where we have assumed that $\omega_{D}^{-1}\approx 0$.

\section{\protect\bigskip CONCLUSIONS}

We have performed a completely original analysis of the equilibrium thermodynamics of a photon gas in the near-field. Through our approach we obtain the near-field entropy which permits the estimation of the entropic-forces contribution to the Casimir effect. The crucial point here has been the elimination of the low frequency modes through the introduction of a cut-off frequency which takes into account confinement effects. At zero temperature, we obtain the well-known $l^{-4}$ behavior of the pressure which corresponds to the interaction between two perfectly conducting plates as obtained by H.B.G. Casimir and at finite temperature we find Lifshitz's results \cite{Landau80a}.

In the nonequilibrium case, our approach enables us to compute the entropy production in the relaxation processes which the interaction between both materials at different temperatures brings about. The physics behind this is that in order to maintain the stationary state in the system, the excess of heat in each one of the bodies must be dissipated into the respective baths since it is a well-known fact that dissipation is concomitant with relaxation.

It must also be stressed that since the thermal frequency $\omega_0$ is proportional to the temperature, $\tau_{\alpha}^{-1}\left(\omega,T_{\alpha}\right) $ vanishes at $T_{\alpha }=0$ and only relaxation processes related to Einstein's stimulated and spontaneous emission remain \cite{Einstein17,Lewis73}. These processes do not depend on temperature and thus, are related to the so-called vacuum fluctuations. Therefore, to account for nonequilibrium Casimir forces, the contribution from vacuum fluctuations must be added to thermal effects.

As a summary, our paper provides a sound starting point for further studies in the field of nanoscale energy exchange with a high potential for technological applications.

\begin{acknowledgement}
This work has been supported by MINECO of the Spanish Government and Consejo Superior de Investigaciones Científicas under Grant No. FIS2015-67837-P, CNPq of the Brazilian Government under Grant No. 405319/2016-9.
\end{acknowledgement}

\end{document}